\RequirePackage{rotating}

\documentclass[twocolumn]{aastex631}
\usepackage{graphicx}
\usepackage{epsfig}
\usepackage{natbib}
\usepackage{multirow}
\usepackage{amsmath}
\usepackage{textcomp}
\usepackage{rotating}
\usepackage{graphicx}
\usepackage{comment}

\newcommand{\lya}{Ly$\alpha$}

\newcommand{\hi}{H\thinspace{\sc i}}

\newcommand{\nv}{N\thinspace{\sc v}}

\begin{document}

\title{The Pitfalls of Using Lyman Alpha Damping Wings in High-z Galaxy Spectra to Measure the Intergalactic Neutral Hydrogen Fraction}
\correspondingauthor{Mason Huberty}
\email{huber458@umn.edu}
\author[0009-0002-9932-4461]{Mason Huberty}
\affiliation{Minnesota Institute for Astrophysics, University of Minnesota, 116 Church Street SE, Minneapolis, MN 55455, USA}

\author[0000-0002-9136-8876]{Claudia Scarlata}
\affiliation{Minnesota Institute for Astrophysics, University of Minnesota, 116 Church Street SE, Minneapolis, MN 55455, USA}

\author[0000-0001-8587-218X]{Matthew J. Hayes}
\affiliation{Stockholm University, Department of Astronomy and Oskar Klein Centre for Cosmoparticle Physics, AlbaNova University Centre, SE-10691, Stockholm, Sweden.}

\author[0000-0002-5659-4974]{Simon Gazagnes}
\affiliation{Department of Astronomy, The University of Texas at Austin, 2515 Speedway, Stop C1400, Austin, TX 78712, USA}

\begin{abstract}
The Lyman Alpha damping wing is seen in absorption against the spectra of high-redshift galaxies. Modeling of this wing is a way to measure the volume neutral hydrogen fraction of the Universe, as well as to measure the strength of \hi\ absorption in the form of a damped \lya\ absorber (DLA). We use very high-quality ultraviolet spectra of low-z galaxies to create mock high-redshift, low-resolution spectra complete with a damping wing resulting from the IGM and \hi\ absorption from local, dense gas to mimic current James Webb Space Telescope (JWST) observations of the early universe. The simulated spectra are used to test the ability to recover the ``true" values of the column density of the DLA ($N_{\rm DLA}$) and the volume neutral hydrogen fraction ($x_{\rm HI}$) of the IGM through modified stellar continuum fitting. We find that the ability to recover the true values of $N_{\rm DLA}$ and $x_{\rm HI}$ simultaneously is compromised by degeneracies between these two parameters, and the spectral shape itself at low-resolution (R$\sim$100). The root mean squared error between the recovered and true column density of the DLA is on the order of 1 dex. This error somewhat decreases with improved resolution (R$\sim$1000), but systematically underestimates $N_{\rm DLA}$ when \lya\ emission is present in the mock spectra. The recovered values of $x_{\rm HI}$ are poorly constrained and do not improve substantially with the higher resolution. We recommend accounting for these sources of uncertainty and biases when using galaxies' \lya\ damping wings to measure the intergalactic neutral hydrogen fraction.

\end{abstract}

\section{Introduction}\label{section1}
The Epoch of Reionization (EoR) marks the last major phase transition in the Universe, during which Lyman-continuum emitting sources ionized the neutral hydrogen in the diffuse intergalactic medium (IGM). Despite the major advancements in the knowledge of the galaxy and active galactic nuclei (AGN) populations at redshift $z\gtrsim 6$ enabled by revolutionary James Webb Space Telescope (JWST) data, many aspects of the reionization epoch are still uncertain. For example, the sources responsible for the production of the bulk of the necessary ionizing radiation are still debated, with galaxies and AGN contending this role \citep[e.g.,][]{madau2024,atek2024}. Additionally, the time and spatial progression of reionization are still almost completely unconstrained \citep[see][for a review of reionization]{robertson2022,finkelstein2024}.

The reionization history of the Universe can be characterized by the volume average neutral hydrogen fraction ($x_{\rm HI}$) as a function of redshift. Among various techniques, $x_{\rm HI}$ can be estimated by observing the red \lya\ damping wings imprinted in quasar (QSO) spectra due to absorption of \lya\ photons by neutral hydrogen in the diffuse IGM. This technique  has been employed for decades \citep[e.g.,][]{fan2006}, enabled by high resolution spectrographs on 10m-class telescopes  \citep[e.g.,][]{murphy2003}. QSOs are the perfect targets for such studies, being intrinsically very bright ($M_{UV}<-27$) and visible out to the highest redshifts. An $M_{UV}=-27$ QSO at $z=6$ corresponds to an observed magnitude of $\sim21$ in the J band, a value within reach of ground based telescopes \citep[e.g.,][]{dodorico2023}. 
AGN are likely the most optimal sources to study $x_{\rm HI}$, because they have ionized all the local \hi\ \citep[][]{hennawi2024}.
These studies find that $x_{\rm HI}$ increases with redshift from $x_{\rm HI}=0$ at $z\sim5-6$ \citep[e.g.,][]{bosman2022}.

Recently, thanks to the unprecedented infrared sensitivity of JWST data, we have been able to begin pushing this technique to much fainter sources, and various studies have proposed the use of galaxies (thousands of times more common than bright QSOs) to measure $x_{\rm HI}$ \citep{umeda2023}.
However, using star-forming galaxies for this purpose introduces several additional challenges. First, since galaxies are typically much fainter than QSOs, high spectral resolution observations become quickly time prohibitive, even with JWST.
Second, unlike QSOs, the intrinsic spectra of galaxies show large variability due to the significant variation in their physical environments (e.g., dust content, stellar ages, nebular continuum, and emission line strength). 
Finally, galaxies in the early Universe are predicted to harbor large reservoirs of neutral hydrogen that will produce strong \lya\ absorption, whose shape, commonly parametrized with a Voigt profile, will depend, among other things, on the gas column density ($N_{\rm HI}$)
\citep{heintz2024,umeda2023,hainline2024,carniani2024}.
For gas with $N_{HI}$ above $2\times 10^{20}$ atoms cm$^{-2}$ (known as damped \lya\ absorbers, DLAs), the damping wings are significant, becoming degenerate with the effect of  $x_{\rm HI}$ (see Figure~\ref{fig:dlavsigm}). 
These challenges must be carefully considered when using star-forming galaxies to probe the $x_{\rm HI}$ history of the Universe.

In this study, we use high resolution spectra of $z<0.1$ galaxies with known reservoirs of neutral hydrogen to simulate mock low-resolution spectra of galaxies during the EoR. 
We use these simulated spectra to quantify the uncertainties and the degeneracies in the estimate of $x_{\rm HI}$ introduced by typical observational limitations.

The paper is organized as follows. In Section~\ref{sec:data} we present the low-redshift spectra and their subsequent manipulation. In Section~\ref{sec:fitting} we summarize the procedure used to recover the IGM and galaxies' properties. Section~\ref{sec:results} presents and discusses the results. In Section~\ref{sec:conclusion} we outline our main conclusions. We assume a flat $\Lambda$CDM cosmology ($h = 0.69$, $\Omega_B=0.047$, and $\Omega_M=0.32$) \citep{planck2020} and use the AB magnitude system \citep{oke1983}.

\section{Data}\label{sec:data}
In order to test how well the universe neutral hydrogen fraction ($x_{\rm HI}$) and the host galaxy's hydrogen column density (hereafter, $N_{\rm DLA}$) can be recovered from the spectra of high redshift galaxies, we require spectra for which $x_{\rm HI}$, $N_{\rm DLA}$, and the galaxy's physical properties are well known.  
Here, we make use of the extensively studied UV spectra of the COS Legacy Archive Spectroscopy SurveY \citep[CLASSY, as presented in][]{berg2022,james2022}, which contains 
45 star-forming galaxies. These galaxies were selected to span a range of star-formation rates (SFR), stellar masses, and metallicities, representative of galaxies at $z\approx 3$ \citep{mingozzi2024,hayessaldana2024}. 
The metallicities of the CLASSY galaxies are on the order of some of the furthest EoR galaxies \citep[$12+\rm log(O/H)\sim7-8$, e.g.,][]{williams2023,langeroodi2023}.
The CLASSY data include complete FUV spectral coverage at resolution $R\sim 5000$, in the wavelength range between $\sim1200$\AA\ to $\sim2000$\AA\ for each galaxy, enabling the measurement of their \hi\ content and column density ($N_{HI}$).

We limit the amount of CLASSY galaxies that can be used in this study: spectra at a redshift of $z\gtrsim0.1$ do not have enough data points red-ward of the damping wing to fit the spectral shape when binned (as described in the next section).
Therefore we apply a redshift cut at $z<0.1$, resulting in 34 CLASSY galaxies that can be used in these simulations.

At $z\lesssim 0.02$, the \hi\ absorption lines can be affected by the Milky Way's \hi\ absorption. 
\citet{hu2023} model the Milky Way \hi\ absorption with a Voigt profile Galactic \hi\ column density derived from 21 cm emission in conjunction with the host galaxy's DLA, so that \hi\ absorption associated with the Milky Way can be identified. 
We use the CLASSY spectra with the Milky Way \hi\ absorption Voigt profiles corrected, so that any \hi\ absorption can primarily be attributed to that of the host galaxy. 

\citet{hu2023} focused on the analysis of the \lya\ properties of the CLASSY galaxies, and demonstrated the variety of shapes that the profile can have, from pure nebular \lya\ emission to damped \lya\ absorption due to the presence of a large reservoir of \hi, and sometimes combination of the two. They identified damped \lya\ absorbers (DLAs, $N_{HI}>2\times10^{20}$) in the spectra of 31 galaxies, for which the column densities were derived by fitting the spectrum around \lya\ with a combination of a stellar population synthesis model plus a modified Voigt profile. The Voigt profile was computed assuming a Doppler parameter of $30\hspace{0.1cm}\rm km\hspace{0.1cm} s^{-1}$ and was allowed to have a velocity offset (with respect to the host galaxy's observed redshift) and a velocity-independent covering fraction.
CLASSY galaxies J0127-0619, J1044+0353, J1105+4444, and J1359+5726 were best fit with two absorbers and thus have two corresponding measures of $N_{\rm DLA}$ in \citet{hu2023}; we take the representative $N_{\rm DLA}$ to be the combined column densities of both absorbers (which is generally very close to the value of the Voigt profile with the larger column density). 

There are also 14 CLASSY galaxies that do not have significant \hi\ absorption detected in the spectra. We do not exclude these objects from the analysis. \citet{hu2023} demonstrated that the presence of strong \hi\ absorption is more common among the galaxies at lower redshift, strongly suggesting that the visibility of the absorption depends on the relative size of the spectroscopic aperture with respect to the \lya\ halo. With JWST spectroscopic observations performed using microshutters, the unknown extent of the ubiquitous \lya\ halos introduces a large source of uncertainty in the intrinsic galaxy spectrum.

In the following, we analyze CLASSY spectra both with and without a measurable DLA.

\subsection{Simulations}\label{sec:simulation}
The goal of this paper is to study how well the properties of the IGM, the galaxy, and its \hi\ content can be recovered from the low-resolution JWST spectra of galaxies well into the reionization epoch. We take as a ``posterchild" example the galaxy CEERS-43833 \citep[see][]{finkelstein2023,arrabal2023}, a spectroscopically confirmed $z=8.76$. 
This object is also one of the first $z\gtrsim 8$ galaxies for which a strong DLA has been identified thanks to excess \hi\ absorption compared what is expected by the diffuse IGM \citep[][]{heintz2024}. This galaxy was observed with the NIRSpec prism, at a resolution of $R\sim100$. 
We used the CLASSY spectra to generate spectroscopic observations of high redshift galaxies tuned to reproduce the resolution of the CEERS-43833 data. 

The steps involved in the simulation of the high redshift spectra are as follows: first, we redshift the CLASSY spectra to $z_{source}=8.76$.
We then apply the IGM attenuation assuming a true neutral hydrogen fraction, $x^T_{\rm HI}$.
We choose three different true values of $x^T_{\rm HI}: 0.1, 0.5,$ and $1.0$. The attenuation on the galaxy's profile is characterized by the IGM optical depth \citep{miralda1998,madau2000,fan2006,totani2006,umeda2023} which can be computed as follows, as a function of $x^T_{\rm HI}$:

\begin{figure}
    \centering
    \includegraphics[width=.9\linewidth]{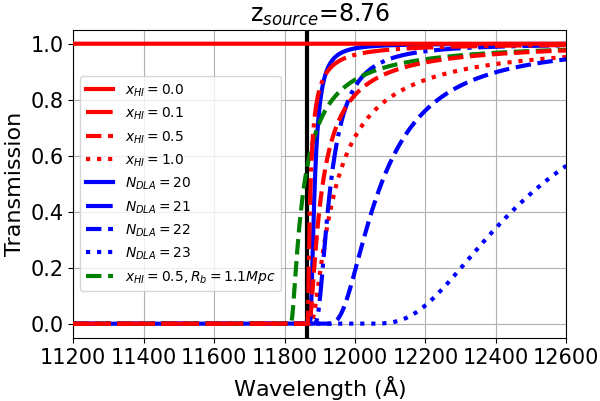}
    \caption{The effects on transmission for varying levels of $x_{\rm HI}$ (red lines) and $N_{\rm DLA}$ (blue lines, and only showing the red-side of the Voigt profiles) for a galaxy at redshift z=8.76. Notice the possibility for misinterpretation of the red damping wing, for example a system with $x_{\rm HI}$=0.1 with no DLA could be easily interpreted as a a system with $x_{\rm HI}=0.0$ and $N_{DLA}=20$. This may lead to extensive degeneracies when fitting both simultaneously, in particular in low-resolution data. The green line shows the IGM transmission for $x_{\rm HI}=0.5$ but with an added bubble of ionized hydrogen surrounding the host galaxy, with a radius of $1.1$ physical Mpc.}
    \label{fig:dlavsigm}
\end{figure}

\begin{figure*}
    \centering
    \includegraphics[width=.9\linewidth]{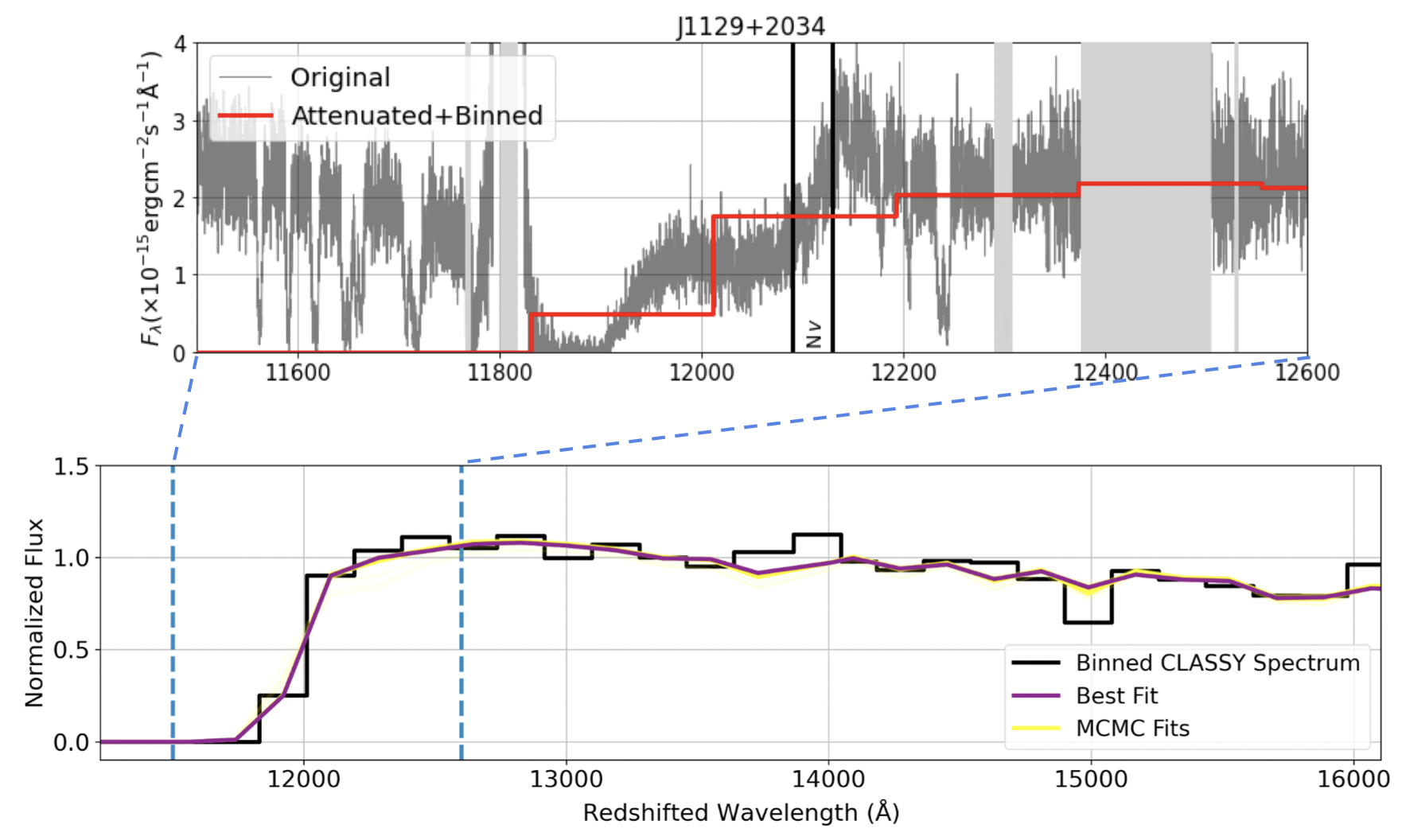}
    \caption{\textbf{Top}: An example of a redshifted, binned, IGM attenuated (with $x_{\rm HI}=0.1$) mock spectrum (red) alongside its original CLASSY counterpart (dark gray). Light gray regions indicate a region that is masked, either from missing data or a Milky Way line. Notice how the DLA, the IGM damping wing, and the \nv\ lines of the SED are all blended together in the binned data. \textbf{Bottom}: An example fit to the mock spectra for J1119+5130 with a neutral hydrogen fraction of $x_{\rm HI}=0.5$ (black, which is the normalized version of the red line in the top panel). The best fit is in purple, and 100 randomly sampled MCMC chains from the posterior distributions highlighted in yellow.}
    \label{fig:binnedexample}
\end{figure*}

\begin{equation}\label{eq:finaligm}
\begin{split}
\tau_{IGM}(z_{obs})&=\tau_{GP}(z_{source})\frac{R_\alpha}{\pi}\left(\frac{1+z_{obs}}{1+z_{source}}\right)^{3/2}\\
&\times \left[I\left(\frac{1+z_{bubble}}{1+z_{obs}}\right)-I\left(\frac{1+z_{end}}{1+z_{obs}}\right)\right]
\end{split}
\end{equation}

\noindent 
where $R_\alpha=\Lambda_\alpha\lambda_\alpha/(4\pi c)=2.02\times10^{-8}$, 
$I(x)=\frac{x^{9/2}}{1-x}+\frac{9}{7}x^{7/2}+\frac{9}{5}x^{5/2}+3x^{3/2}+9x^{1/2}-\frac{9}{2}ln\left[\frac{1+x^{1/2}}{1-x^{1/2}}\right]$, and 

\begin{equation}\label{eq:fan}
\begin{split}
\tau_{GP}(z)&=1.8\times 10^5 h^{-1}  \Omega_M^{-1/2} \frac{\Omega_B h^2}{0.02} \left(\frac{1+z}{7}\right)^{3/2}x^T_{\rm HI}\\
&\approx 2.79 \times 10^4 (1+z)^{3/2} x^T_{\rm HI}
\end{split}
\end{equation}

\noindent 
and we use $h = 0.69$, the baryonic density parameter $\Omega_B=0.047$, and a matter density parameter $\Omega_M=0.32$  \citep{planck2020}. 
Additionally, $z_{obs}=\frac{\lambda_{obs}}{\lambda_\alpha}-1$, $z_{source}$ is the redshift of the source galaxy, $z_{end}$ is the redshift marking the end of reionization (which we approximate as $z = 6$), and $z_{bubble}$ is the redshift corresponding to the edge of the ionized bubble surrounding the galaxy. For simplicity, we initially assume that there is no substantial ionized bubble around the galaxies, i.e., $z_{bubble}=z_{source}$. However we note that in reality bubbles are expected and observed \citep{hayesscarlata2023}. We explore the effects of the addition of a bubble in Section~\ref{sec:bubble}.

We mask Milky Way absorption lines detected in the CLASSY spectra, which would not be present in true high redshift spectra (during the binning process these points are not included in the bin). 
Next, we bin the spectra to the resolution of the NIRSpec prism, $R=\frac{\lambda}{\Delta \lambda}\approx100$ \citep[see][]{jakobsen2022}.

Typical CLASSY spectra have a continuum signal-to-noise ratio of $\gtrsim5$ per $100 \rm km s^{-1}$ resolution element, which increases to $\gtrsim80$ after binning. This SNR is higher than for typical NIRSpec prism spectra \citep[e.g., $\gtrsim 3$ for CEERS-43833), see][]{heintz2024}. Therefore, we can treat the binned CLASSY data as a ``best-case scenario" in regards to signal-to-noise.

An example outcome of the binning process for galaxy J1129$+$2034 can be seen in the top panel of Figure~\ref{fig:binnedexample}. From the analysis of the COS spectra, this galaxy is known to have a young stellar age of $2.56\pm0.79$ Myrs, and a strong DLA with a column density of log$(21.11\pm0.01 /\rm cm^{-2})$, about an order of magnitude lower than what was found in CEERS-43833). 
We redshift all 34 CLASSY spectra, resulting in a sample of 34$\times$3 (one for each of the three values of $x_{\rm HI}$) ``mock" low-resolution spectra of high-redshift galaxies with known $x_{\rm HI}$ and $N_{\rm DLA}$ \citep[from the analysis of][]{hu2023}. We use these spectra in the next section to explore the systematic uncertainties in the recovery of the galaxies' and IGM's properties.  

\section{Fitting Procedure}\label{sec:fitting}
We use a modified version of the \textit{FItting the stellar Continuum of Uv Spectra} (\textit{FICUS}) \textit{Python} code \citep{saldana2023}, adapted to simultaneously fit the stellar continuum, alongside the properties of a DLA and the IGM. 
\textit{FICUS} uses stellar population synthesis models computed using \textit{Starburst99} \citep{leitherer2011,leitherer2014}, 
a Kroupa initial mass function \citep{kroupa2001}, varying dust content computed using the attenuation curves of \citet[][]{reddy2016},  
and a nebular continuum and emission lines generated through self-consistently processing the stellar population models using the \textit{CLOUDY} photoionization code \citep{ferland2017}. \textit{FICUS} 
fits a linear-combination of stellar population templates to the observed data, at a specified spectral resolution. 

We modified FICUS to allow for two new components in addition to the stellar templates: a module describing the absorption profile of the DLA and a module one accounting for the IGM absorption. We use Equation~\ref{eq:finaligm} to describe the red damping wing of the IGM absorption, with only the neutral hydrogen fraction ($x_{\rm HI}$) as a free parameter. The DLA absorption is parameterized using a Voigt profile \citep[e.g.,][]{tepper2006, prochaska2015}, which depends on the optical depth due of \lya\ photons, described as

\begin{equation}
\tau_{DLA}=N_{DLA}C a H(a,x)
\end{equation}

\noindent 
where $C$ is the photon absorption constant, $a$ is the damping parameter (dependent on the Doppler parameter $b$), and $H(a,x)$ is the Voigt-Hjerting function. We follow \citet{prochaska2015} and \citet{hu2023}, and assume a Doppler parameter of $b = 30 \rm km s^{-1}$. For \lya, we can fix $b$ because it has a relatively small impact on the inferred $N_{\rm DLA}$ value. 
These modifications result in two additional variables that control the shape of the spectrum around the \lya\ wavelength: $x_{\rm HI}$ and $N_{\rm DLA}$. In Figure~\ref{fig:dlavsigm} we illustrate how these two variables are highly degenerate with each other.
For example the profile with $x_{\rm HI}=0.1$ and no \hi\ reservoir is very similar to the profile with $log_{10}N_{DLA}/{\rm cm}^{-2}=20$ and no IGM absorption. The small visible differences would disappear when low-resolution data are considered. To try and minimize this effect, some studies limit the detection of $N_{\rm DLA}$ to values larger than  $log_{10}N_{DLA}/{\rm cm}^{-2}=21$, to represent cases where the damping wing can be primarily attributed to local \hi\ gas \citep{primal2024}.

We use the \textit{emcee} Monte-Carlo Markov Chain (MCMC) \citep{foreman2013} code within modified FICUS to find the best fit stellar population parameters, dust extinction, $x_{\rm HI}$ and $N_{\rm DLA}$. 
In the fitting procedure, log$_{10}N_{DLA}$ is constrained between 18 and 22.5, $x_{\rm HI}$ between 0 and 1, and $E_{b-v}$ between 0 and 0.5 assuming a \citet{reddy2016} dust attenuation curve. This law is typically assumed for UV stellar continuum fitting \citep[e.g.,][]{saldana2023}.

\begin{figure*}
    \centering
    \includegraphics[width=1.0\linewidth]{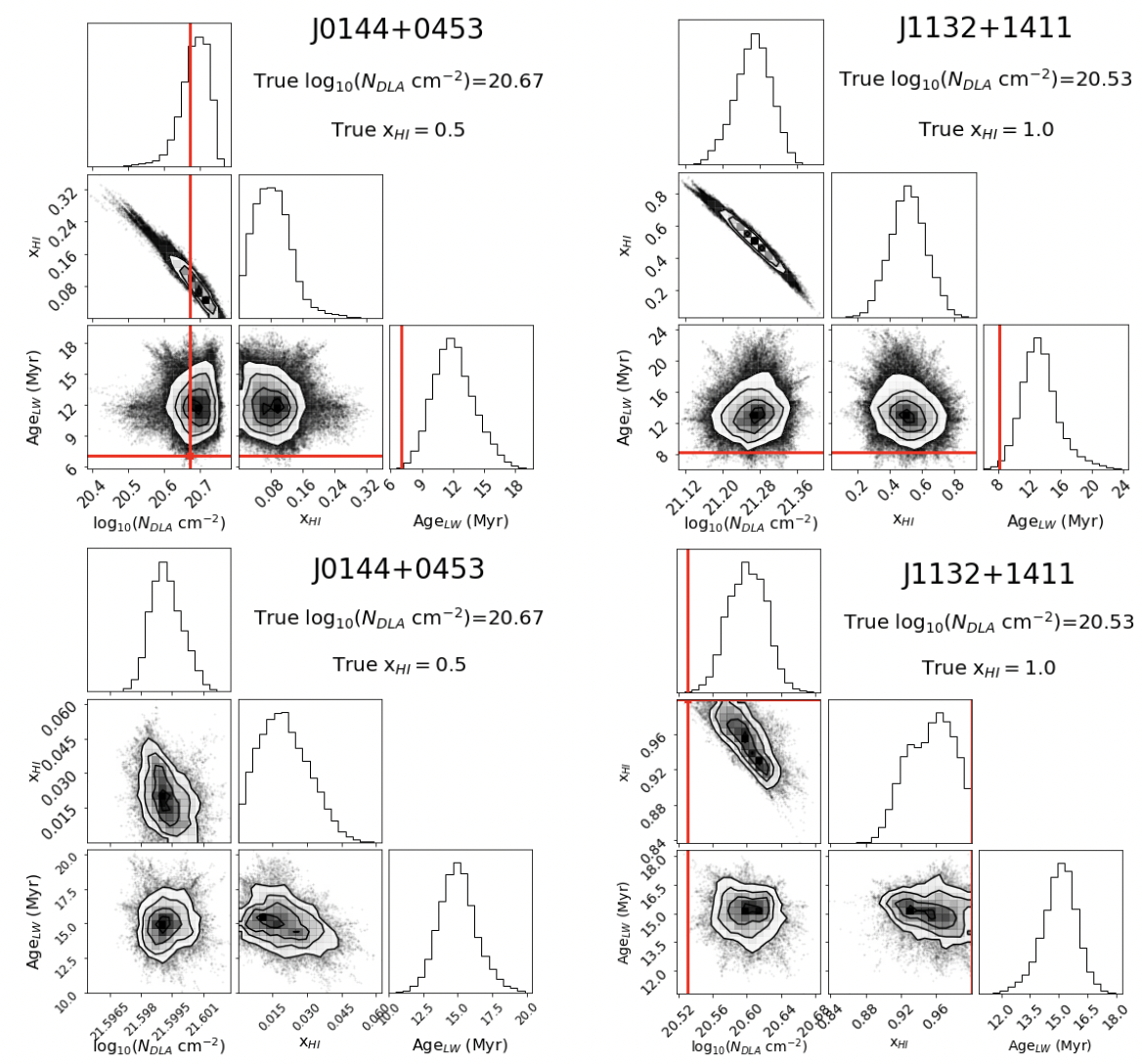}
    \caption{Example corner plots demonstrating the degeneracies in the $x_{\rm HI}$, $N_{\rm DLA}$, and $Age_{LW}$ (luminosity-weighted age) parameter space of the damping wing. \textbf{Top row}: Corner plots for spectra with $R\sim100$. \textbf{Bottom row}: Corner plots for spectra with $R\sim1000$. The true values of the parameters are marked in red (in many cases, the true value is not visible as the MCMC chains find the best fit in another part of the parameter space). A strong degeneracy exists between $x_{\rm HI}$ and $N_{\rm DLA}$, and a lighter one with Age$_{LW}$ which characterizes the spectral shape and the strength of spectral features. The strength of the $x_{\rm HI}$/$N_{\rm DLA}$ is softened with higher resolution.}
    \label{fig:cornersall}
\end{figure*}

We run \textit{emcee} for $N=10000$ MCMC steps, and take the range corresponding to the 16th and 84th percentiles of the posterior chains as the credibility interval.
An example is shown in the bottom panel of Figure~\ref{fig:binnedexample}. Example corner plots are shown in the top row of Figure~\ref{fig:cornersall}.

\section{Results and Discussion}\label{sec:results}
\begin{figure*}
    \centering
    \includegraphics[width=1.0\linewidth]{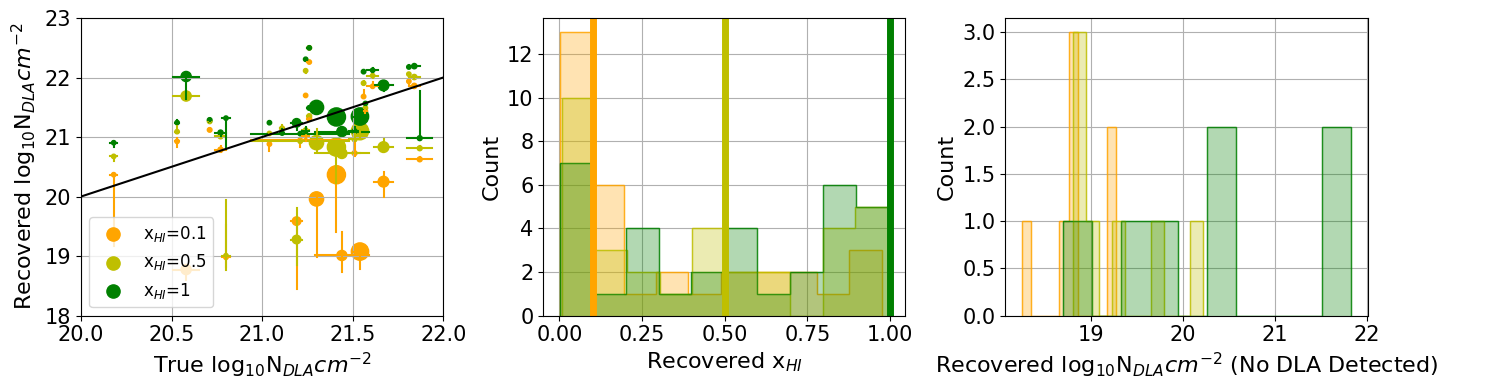}
    \caption{\textbf{Left:} A comparison of the true and recovered values of $N_{\rm DLA}$ in the CLASSY galaxies, for three different values of $x^T_{\rm HI}$. The black line represents when the recovered and true values of $N_{\rm DLA}$ are equivalent. $N_{\rm DLA}$ is more likely to be under-predicted when $x_{\rm HI}=0.1$. The size of the points in the has been scaled to the equivalent width of \lya\ emission present in the intrinsic spectrum.\textbf{Middle:} Histograms of the recovered values of $x_{\rm HI}$ for the three different values of $x^T_{\rm HI}$. The solid vertical lines define the three 'true' values of $x^T_{\rm HI}=0.1, 0.5,$ and $1.0$. The recovered values for $x_{\rm HI}=0.1$ and $1.0$ peak near the true values, but the recovered values span the entire parameter space. 
    \textbf{Right:} Additional predictions for $N_{\rm DLA}$ in spectra where there was no detected DLA in the original data, emphasizing how a DLA can be predicted when there should have been minimal to no prediction.} 
    \label{fig:FICUSnx}
\end{figure*}

We now discuss the results of the simulations and the implications they have for studies of the high-z universe. 
The results are summarized in Figure~\ref{fig:FICUSnx}, where we focus on the recovery of the DLA column density, $N_{\rm DLA}$, and the neutral hydrogen fraction of the IGM, $x_{\rm HI}$, in the mock spectra. We use different colors to identify the three intrinsic values of $x^T_{\rm HI}$ used to simulate the spectra. 
The left panel of Figure~\ref{fig:FICUSnx} shows that the recovered values of $N_{\rm DLA}$ are not necessarily indicative of the ``true", intrinsic values measured in the original, non-IGM attenuated, high resolution data  (Note that we are using ``true" to describe $N_{\rm DLA}$ measured at the highest resolution provided by the CLASSY data). The root mean squared error (RMSE) between the observed and recovered values of $N_{\rm DLA}$ is 1.07, 0.76, and 0.58 dex for $x^T_{\rm HI}=0.1, 0.5,$ and $1.0$ respectively. This indicates that $N_{\rm DLA}$ tends to be recovered more precisely for higher values of $x^T_{\rm HI}$. 
On average, the column density is overestimated by 0.27 and underestimated by 0.53 dex for $x^T_{\rm HI}=1.0$ and $0.1$ respectively and the difference between the true and recovered values can be as high as 3 dex in some of the most extreme cases.

The middle panel of Figure~\ref{fig:FICUSnx} shows that the recovered value of $x_{\rm HI}$ is unconstrained. The recovered values span especially the entire allowed range, with no obvious correlation with the true values. On the one hand, the recovery of $x^T_{\rm HI}=0.1$ does peak near $x^T_{\rm HI}=0.1$. On the other hand, we just showed that the recovered value for $N_{\rm DLA}$ for $x^T_{\rm HI}=0.1$ was the worst of the three values of $x^T_{\rm HI}$. 
The RMSE for the recovered $x_{\rm HI}$ is 0.39, 0.36, and 0.59 for $x^T_{\rm HI}=0.1, 0.5, 1.0$ respectively. 

Finally, the right panel of Figure~\ref{fig:FICUSnx} shows the histograms of the recovered DLA column densities for galaxies that had no detectable DLA in the original CLASSY data (i.e., there is no ``true" value of $N_{\rm DLA}$ for these galaxies). Despite the fact that a DLA is erroneously detected in the simulated spectra, the predicted column densities are, on average, low (with an average of $\approx 10^{19}$cm$^{-2}$). We can also, however, see that there are a few outliers with a much larger predicted column density, mostly associated with large values of $x_{\rm HI}$. Although the signal is not strong, we find that there is an indication of the peak of the recovered $N_{\rm DLA}$ distributions to move to larger values as $x_{\rm HI}$ increases. This is further discussed below.  Overall, our results show significant under and over-estimations of the host galaxy \hi\ content, which is a major concern for reionization-focused studies.

It is important to understand the origin of the discrepancies between the recovered and intrinsic values of $N_{\rm DLA}$ and $x_{\rm HI}$ to identify, if possible, mitigating strategies for future observations. To this aim, 
we start by explore the recovered properties of the stellar population fitting. We use as true values of the luminosity-weighted stellar age and luminosity-weighted metallicity those obtained from the fit of the original, high-resolution CLASSY spectra. The stellar populations were fit using Starburst99 stellar population synthesis models, and the details of the fits can be found in \citet{berg2022} and \citet{hu2023}. It is important to note that we do not use the same assumptions in the spectral modeling as those in  \citet{berg2022}  \citep[done with BEAGLE, see][]{chevallard2016}, such as the assumed initial mass function.

\begin{figure*}
    \centering
    \includegraphics[width=.7\linewidth]{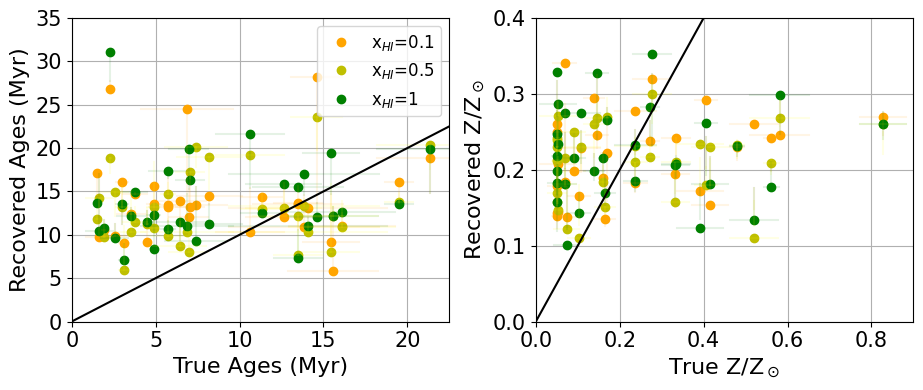}
    \caption{Comparisons between the recovered and true parameters of  stellar age (left), and $Z$ (right). The black lines are the one-to-one lines. The recovered age and metallicity values are quite stagnant across the parameter space. The error-bars are faint to improve readability.}
    \label{fig:FICUSebv}
\end{figure*}

Figure~\ref{fig:FICUSebv} displays the comparison between the true and recovered values of stellar age and metallicity. We find no obvious trends between the recovered luminosity-weighted stellar age and metallicity and their true values. This result is not completely unexpected. Apart from the underlying differences in SED modeling assumptions, the low spectral-resolution of the data ``washes out" age-sensitive stellar-wind features in the spectra \citep[e.g.,][]{leitherer2010,wofford2011,chisholm2019}. Additionally, it causes the blending of the \hi\ damping wings with the \nv\ stellar absorption present in some of the youngest starbursts.
We can see an example of the former effect in the bottom panel of Figure~\ref{fig:binnedexample}, where it is clear that the CIV stellar features at 1550\AA, which are reliable tracers of age and metallicity, are only seen in a single data point. 
Hence, only the slope of the continuum can be used as a constraint on the stellar population age, but this is highly degenerate with the dust attenuation as well \citep{chisholm2022} and strength of the nebular continuum emission.

The impact of the \lya/\nv\ blending can clearly be seen in the top panel of Figure~\ref{fig:binnedexample}. Because of its proximity to \lya\ the presence of strong \nv\ $\sim1240$\AA\ absorption can mistakenly be interpreted as due to the wings of the \hi\ absorption (either in the DLA or in the IGM). For a $N_{DLA}= 10^{20}$ cm$^{-2}$ DLA, the stellar continuum at the wavelength of \nv\ is suppressed by $\approx 40$\%, similar to the impact of the \nv\ stellar absorption. This degeneracy makes it difficult to separate these two components, particularly in the youngest galaxies ($\lesssim5$ Myr) where the \nv\ is strongest.

To explore degeneracies among the recovered parameters, we examine MCMC corner plots obtained from the fitting procedure. Examples can be seen in the top row of Figure~\ref{fig:cornersall}. 
The most striking degeneracy lies in the relationship between $N_{\rm DLA}$ and $x_{\rm HI}$. A strong correlation exists in the joint-distribution between these two parameters, as $N_{\rm DLA}$ increases, $x_{\rm HI}$ decreases and vice-versa, as expected. For these two examples, the Pearson correlation coefficients between $N_{\rm DLA}$ and $x_{\rm HI}$ are -0.96 and -0.98 for J0144+0453 and J1132+1411 respectively. 

A possible degeneracy also exists between $N_{\rm DLA}$/$x_{\rm HI}$ and the luminosity-weighted age.
As $N_{\rm DLA}$ increases, the age tends to increase as well (though much weaker than the $N_{\rm DLA}$/$x_{\rm HI}$ relation), showing how you can reproduce the same spectra with either low $N_{\rm DLA}$ and young ages, or high $N_{\rm DLA}$ and old ages. Therefore, the correlation between recovered age and $x_{\rm HI}$ noted above, may be the result of a $N_{\rm DLA}$/$x_{\rm HI}$ degeneracy: larger $N_{\rm DLA}$ correlates with lower $x_{\rm HI}$ and higher age.

\begin{figure*}
    \centering
    \includegraphics[width=1.0\linewidth]{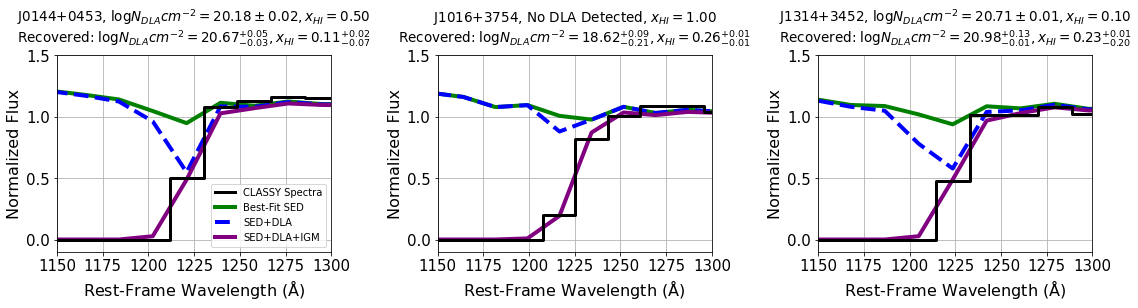}
    \caption{Three examples of \lya\ break fits, showing how the different components of the fit make up the total profile. The spectral shape is in green, the addition of HI absorption is shown in blue, and the addition of the IGM is shown in purple. \textbf{Left:} A fit where $x_{\rm HI}$ is under-predicted at the expense of over-prediction of $N_{\rm DLA}$.
    \textbf{Center:} A fit where $N_{\rm DLA}$ is added to balance the under-estimate of $x_{\rm HI}$. \textbf{Right:} A fit where  $x_{\rm HI}$ is recovered within $1\sigma$ but overestimates $N_{\rm DLA}$ by a more substantial margin. It recovers weaker \nv\ absorption, with the consequence of an increase in the predicted stellar age, and a decrease in the predicted metallicity, and it is this spectral shape that helps compensate for the overestimate in $N_{\rm DLA}$.}
    \label{fig:dlaigmeffects}
\end{figure*}

To further explore the origin of this bias, we consider three galaxies in detail in Figure~\ref{fig:dlaigmeffects}. 
In each panel, we show the low-resolution spectrum with the black line. We also show the individual components that enter the best fit: the green line shows the best fit stellar$+$nebular continuum, the dashed blue line shows the spectrum after the absorption due to the \hi\ reservoir, while the solid purple line shows the best fit total spectrum, including the IGM attenuation.

In the left panel, we show the result for J0144$+$0453, which, according to \citet{hu2023}, has a $\log {(N_{DLA}/ \rm cm^2)}=20.18\pm0.02$. The spectrum shown in the figure was simulated assuming a $x^T_{\rm HI}=0.50$. For this object, the recovered DLA column density is overestimated by about $1/2$ dex. In this case, most of the absorption that is truly due to the IGM, is attributed to the DLA, resulting in a recovered value of $\log {(N_{DLA}/ \rm cm^2)}=20.67^{+0.05}_{-0.03}$.

The center panel shows galaxy J1016$+$3754 which does not show significant DLA in the original spectrum. We simulated the mock spectrum assuming $x_{\rm HI}=1.0$. In this galaxy, it is predicted that the IGM attenuation is much smaller than the true value ($x_{\rm HI}\approx0.26$). However, this is still enough absorption to fit the damping wing without substantial contribution from the DLA, so we underestimate $x_{\rm HI}$ because we overestimate $N_{\rm DLA}$.

Finally, in the right panel, we show J1314$+$3452, an example for which $x_{\rm HI}$ is recovered within $1\sigma$ of the true value, but $N_{\rm DLA}$ is overestimated by $\sim0.3$ dex. J1314$+$3452 is a galaxy with a young stellar population, with the true values of stellar age and metallicity measured to be 2.31$\pm0.04$ Myr and 0.41$\pm0.03$ $Z_{\odot}$. 
The recovered value for these parameters are $16.8^{+0.0}_{-6.0}$ Myr (seven times larger than the true value) and $0.20^{+0.03}_{-0.01}$ $Z_{\odot}$ (two-fold smaller than the true value) respectively, changing the shape of the recovered stellar continuum before any DLA and IGM absorption. This example highlights the importance of the \nv\ spectral feature: as the age of a burst increases, or as the metallicity decreases, the \nv\ becomes less pronounced  \citep{leitherer2010,wofford2011,chisholm2019}. Accordingly in this galaxy the larger recovered DLA column density is needed to compensate for a weaker recovered absorption at \nv.

To summarize, we found that because the \hi\ in the IGM and DLA absorbs over the same spectral range close to \lya\ their effect is degenerate. Additionally, we found that the stellar continuum can contribute to this degeneracy because of strong stellar features in the vicinity of \lya. Our results show that low resolution data cannot effectively separate these components and the recovered parameters suffer from large uncertainties. 

\begin{figure*}
    \centering
    \includegraphics[width=.9\linewidth]{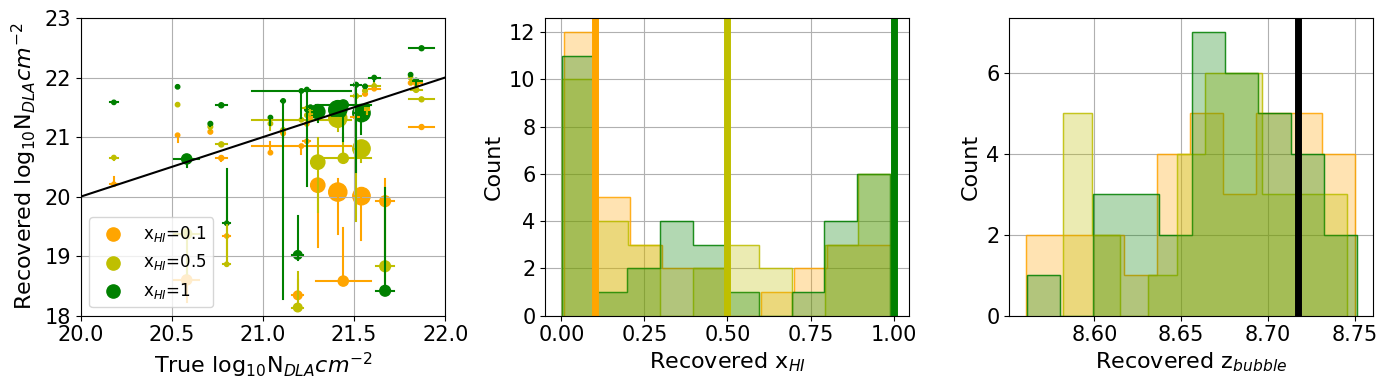}
    \caption{Left and middle panel are the same as Figure~\ref{fig:FICUSnx}, except for mock spectra with a bubble of ionized gas. The right panel is the recovery of $z_{bubble}$ as used in Equation~\ref{eq:finaligm}. In the left panel, the size of the points in the has been scaled to the equivalent width of \lya\ emission present in the intrinsic spectrum. In the right panel, the black vertical line represents the true value of $z_{bubble}$.}
    \label{fig:bubblesnx}
\end{figure*}

\subsection{Fits with an Ionized Bubble}\label{sec:bubble}
Previously, we assumed that there existed no substantial bubble of ionized hydrogen surrounding the source galaxy, or in other words, that $z_{bubble}=z_{source}$ in Equation~\ref{eq:finaligm}. However, it is well established that throughout the EoR, these bubbles are present and are measured to grow increasingly large with lower redshift \citep[e.g.,][]{umeda2023,hayesscarlata2023,mason2025}. 
The effect of the addition of this bubble can be seen in the green line in Figure~\ref{fig:dlavsigm}. The inclusion of a bubble adds a blue ``tail" to the damping wing, allowing for non-zero transmission blueward of the \lya\ wavelength. On the one hand, an additional parameter may add to the uncertainties found with the low resolution mock spectra. On the other hand, the additional blue tail may help remove some of the degeneracies by, in some sense, partially ``shifting" the absorption due to the ISM blue-ward. 

We chose to create a new batch of mock spectra with a bubble radius of $1.1$ physical Mpc, corresponding to $\approx 10.7$ commoving Mpc at the simulated redshift of 8.76. This value was chosen as the median value from the \citet{hayesscarlata2023} sample.
This bubble corresponds to $z_{bubble}\approx 8.72$, as used in Equation~\ref{eq:finaligm}. 
We repeat the simulation and modeling procedure for spectra with $x^T_{\rm HI}=0.1, 0.5$ and $1.0$ as outlined in Section~\ref{sec:simulation} with this $z_{bubble}$ modification. The fitting range for $z_{bubble}$ is set to [8.56,8.76] which corresponds to a bubble size range of about 0 to 50 commoving Mpc. 
The results are seen in Figure~\ref{fig:bubblesnx}. 

The left panel of Figure~\ref{fig:bubblesnx} shows the recovery of $N_{DLA}$ in mock spectra with the added bubble. The size of the data points in this panel is proportional to the equivalent width of the \lya\ emission in the intrinsic spectra as measured in \citet{hu2023}.
The RMSE between the observed and recovered values of $N_{\rm DLA}$ is 1.11, 1.00, and 0.95 dex for $x^T_{\rm HI}=0.1, 0.5,$ and $1.0$ respectively. 
Thus, we see a poorer recovery of $N_{\rm DLA}$ when the bubble was added. 
We also note that the recovered values of N from mock spectra with little to no intrinsic \lya\ emission are generally more accurate than those from spectra with substantial \lya\ emission.

The middle panel shows the recovery of $x_{HI}$. The RMSE for the recovered value of $x_{\rm HI}$ is 0.42, 0.37, and 0.67 for $x^T_{\rm HI}=0.1, 0.5$, and $1.0$ respectively. Thus, we see essentially no improvement in the recovery of $x_{\rm HI}$ with the addition of the bubble in the mock spectra, mirroring the poor recovery of $x_{HI}$ found in simulations without a bubble. 

The right panel shows the recovery of $z_{bubble}$, with the vertical black line corresponding to the assumed true value. The recovery of $z_{bubble}$ does not vary substantially with the intrinsic value of $x^T_{\rm HI}$: the RMSE between the true and recovered values of $z_{bubble}$ are 0.068, 0.067, and 0.061 for $x^T_{\rm HI}=0.1, 0.5, 1.0$ respectively. The average recovered value is 2.24 physical Mpc, i.e.,  the recovered bubbles tend to be overestimated by a factor of about 2. 

In the next section, we explore how the results would change if higher spectral resolution modes available on JWST are used.

\newblock

\subsection{Fits at the resolution of NIRSpec G140M}\label{sec:gratingresolution}
Given the complications of using NIRSpec Prism data identified above, what if instead we were to observe such galaxies with the NIRSpec G140M disperser? G140M probes the 0.97–1.84 micron range, the same range that galaxies of $z\sim8.76$ have a Lyman break, at a resolution of $R\sim1000$. We chose this grism instead of G140H ($R\sim2700$) as a compromise between resolution and exposure time. This resolution is about 10 times larger than that of the Prism, which could alleviate some of the concerns of not having enough information to disentangle the effects of the IGM, \hi\ reservoir, and young stellar population. 

\begin{figure*}
    \centering
    \includegraphics[width=.95\linewidth]{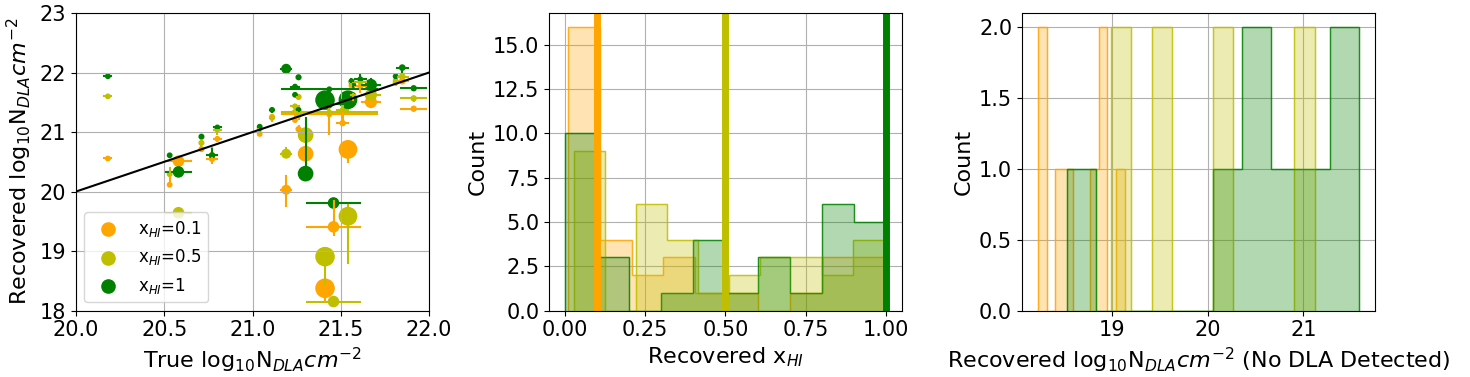}
    \caption{Same as Figure~\ref{fig:FICUSnx}, but using data binned at a resolution of $R\sim1000$ (designed to imitate the resolution of NIRSpec G140M). The size of the points in the left panel is proportional to the equivalent width of the intrinsic \lya\ emission line. This shows that \lya\ emission is causing some of the most severe underestimates of $N_{\rm DLA}$.}
    \label{fig:gratingnx}
\end{figure*}

We return to the original data, and repeat almost the same masking, IGM attenuation, and binning procedure as Section~\ref{sec:simulation} (again with $z_{bubble}=z_{source}$). There are two differences from the original procedure: first, we bin the CLASSY data to $R\sim1000$ instead of $R\sim100$. Second, \lya\ emission is now distinguishable in some of the mock spectra, in which cases it is masked. 
We then repeat the same fitting procedure as described in Section~\ref{sec:fitting}. 
In Figure~\ref{fig:gratingnx}, we display the results. 

The left panel of Figure~\ref{fig:gratingnx} indicates an improvement in the recovery of values of the $N_{\rm DLA}$: the root mean squared errors between the true and recovered value are 0.81, 0.98, and 0.59  for $x^T_{\rm HI}=0.1, 0.5,$ and $1.0$ respectively. This is a relatively large improvement from the Prism fits for the $x^T_{\rm HI}=0.1$ and $1.0$ cases, and it's similar for the $x^T_{\rm HI}=0.5$ case. It remains the case that $N_{\rm DLA}$ is recovered best for higher values of $x_{\rm HI}$. The corner plots also show change: the bottom row of Figure~\ref{fig:cornersall} (which comes from the same mock spectra as the top row in Figure~\ref{fig:cornersall}) still shows a correlation between $x_{\rm HI}$ and $N_{\rm DLA}$ that was present at $R\sim100$, but it is not as tight as before, with the Pearson correlation coefficients dropping to -0.38 and -0.75 for J0144+0453 and J1132+1411 respectively (from -0.86 and -0.98). 
This indicates that these two parameters are less coupled than before, but still remain correlated.

\begin{figure}
    \centering
    \includegraphics[width=.9\linewidth]{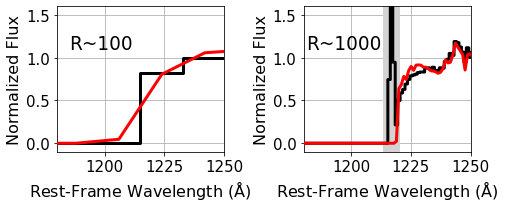}
    \caption{J1359+5726 mock spectra (black) attenuated with $x_{\rm HI}=0.1$ and the subsequent best-fits (red). This galaxy has \lya\ emission that is visible at higher resolution (R$\sim1000$), that can be masked (light gray). However, this line is not visible at lower-resolution (R$\sim100$), and therefore is unmasked. This makes it difficult to constrain $N_{\rm DLA}$ at low resolution.}
    \label{fig:J1359}
\end{figure}

The outliers of $N_{\rm DLA}$ that see the least improvement are those with strong \lya\ emission in the original COS spectra. This is shown in the left panel of the figure, where the size of each point is proportional to the measured equivalent width of \lya\ emission \citep[as measured in][]{hu2023}.

A good example of this is J1359+5726, a galaxy with a true $N_{DLA}=21.44$. This galaxy shows a strong \lya\ emission that clearly affects the shape of the damping wing, which as can be seen in Figure~\ref{fig:J1359}. 
In low-resolution spectra, it is impossible to differentiate host galaxy \lya\ emission with the damping wing (as would be the case for true high-z observations with NIRSpec prism, in the left panel of Figure~\ref{fig:J1359}). At high resolution, we can visually identify \lya\ emission at the resolution of G140M (right panel of Figure~\ref{fig:J1359}) and mask it. 
Unfortunately, having to mask this type of emission in the $R\sim1000$ spectra leads to a loss of data points on the damping wing we seek to model.

In the middle panel of Figure~\ref{fig:gratingnx}, the recovered values of $x_{\rm HI}$ are still not constrained well. The root mean squared error is within 0.02 between the Prism and grating cases for $x^T_{\rm HI}=0.1, 0.5$ (0.39 and 032 respectively) and actually gets slightly larger in the grating case for $x^T_{\rm HI}=1.0$ than for the Prism (up to 0.64).

Another interesting observation is evident in the right plot of Figure~\ref{fig:gratingnx}: when the DLA was not detected in the original CLASSY spectra, the recovered values of $N_{\rm DLA}$ correlate with the true value of $x^T_{\rm HI}$. This was hinted at in the right panel of Figure~\ref{fig:FICUSnx}, but the signal is much stronger with the higher spectral resolution simulations. The higher the value of $x^T_{\rm HI}$, the larger the recovered value of $N_{\rm DLA}$. This strengthens the idea that $N_{\rm DLA}$ and $x_{\rm HI}$ are degenerate.

Thus, while we see some improvement in the prediction of $N_{\rm DLA}$ at higher resolution, there is still much imprecision in the ability to recover DLA and IGM properties, in particular to obtain $x_{\rm HI}$.

\subsubsection{Ionized Bubbles at High Resolution}\label{sec:bubbleshighres}
\begin{figure*}
    \centering
    \includegraphics[width=.9\linewidth]{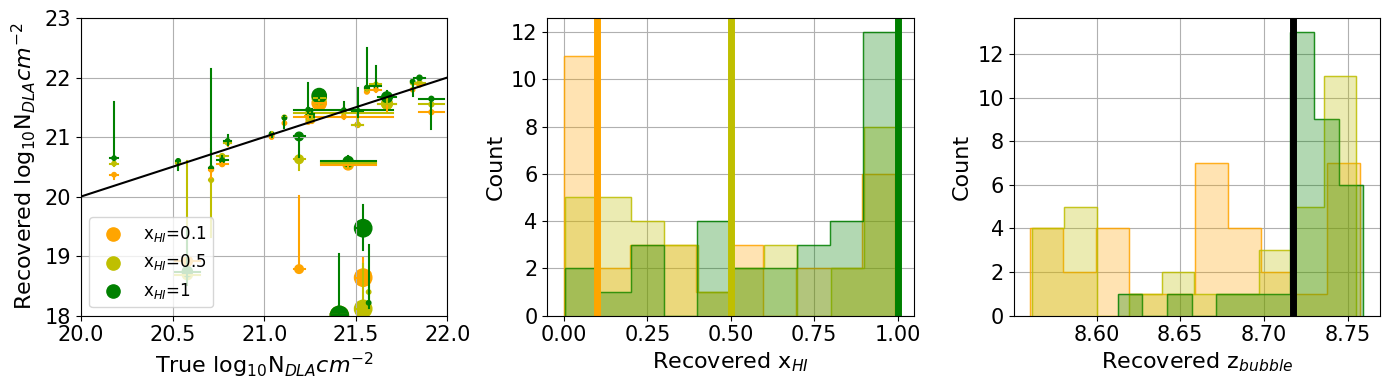}
    \caption{Same as Figure~\ref{fig:bubblesnx}, but for mock spectra with R$\sim$1000.}
    \label{fig:bubbleshighres}
\end{figure*}

We also consider the addition of an ionized bubble as done in  Section~\ref{sec:bubble}, but now for the high resolution mock spectra utilized in Section~\ref{sec:gratingresolution}. The results are displayed in Figure~\ref{fig:bubbleshighres}.

The left panel of Figure~\ref{fig:bubbleshighres} shows the typical strong recovery of $N_{\rm DLA}$ in most galaxies, with the exception of the same few outliers noted in the previous section. The median discrepancy between the true and recovered values of $log_{10}N_{\rm DLA}cm^{-2}$ are 0.03, 0.01, and 0.04 for $x^T_{\rm HI}=0.1, 0.5,$ and $1.0$ respectively, indicating that typically, $N_{\rm DLA}$ is recovered quite well. 

The middle panel shows that when a bubble is present, and high resolution data are used, the recovery of $x_{\rm HI}$ improves somewhat, with RMSEs of 0.47, 0.35, and 0.45 for $x^T_{\rm HI}=0.1, 0.5,$ and $1.0$ respectively. 

The right panel of Figure~\ref{fig:bubbleshighres} shows the recovery of $z_{bubble}$. Mock spectra with intrinsic values of $x^T_{\rm HI}=0.1$ and $0.5$ have RMSEs 0.080 and 0.079 respectively, slightly worse than in the low-resolution mock spectra with ionized bubbles. However, the RMSE of $z_{bubble}$ has dropped substantially in the $x^T_{\rm HI}=1$ case, now at 0.030 (about half the RMSE of the low-resolution version of the simulation). 
With these results, we can conclude that in the case of a fully neutral IGM ($x^T_{\rm HI}=1$), and for galaxies for which a ionized bubble is known to exist, the bubble size can be recovered accurately using high resolution spectra. The universe average ionization fraction, however, remains still largely unconstrained.

\subsection{Other Effects}\label{sec:othereffects}

There are a number of physical effects that we have neglected to model in our simulations that would be present in real observations. 
First, the signal-to-noise of the simulated spectra is much larger than the observed signal-to-noise in real NIRSpec data, making the results optimistic. 

Second, it is possible that the shape of the attenuation law is different in the early universe \citep[e.g.,][]{markov2024}. Typical SED fitting codes use extinction laws known from the local/mid-z universe, whether these apply to the EoR remains to be seen, adding further sources of uncertainties.

Third, while we did consider ionized bubbles using the analytical expression of \citet{miralda1998}, 
\citet{keating2024} found that this simplification does not reproduce the size of ionized bubbles observed in hydrodynamical simulations because of the residual neutral gas within the bubbles themselves  \citep{puchwein2023}.

Fourth, another effect that we did not simulate in our study is the impact of large scale motions of \hi\ gas in the surrounding of the galaxies, reflecting cosmological gas accretion as well as possible large scale outflows driven by stellar feedback. Simulations by Bhagwat et al., (private communication) show that cold streams of \hi\ gas that feed the central galaxies can travel at 200-300 km/s, effectively imprinting a redshift in the absorbing material. Similarly, outflows can travel up to similar velocities, but would imprint a blue-shifted absorption instead \citep[e.g.,][]{huberty2024}. 

\begin{figure}
    \centering
    \includegraphics[width=.9\linewidth]{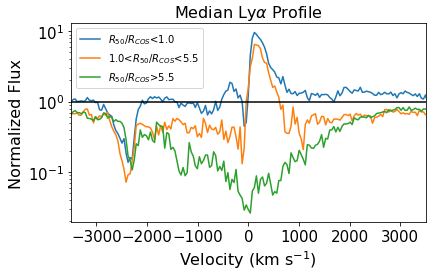}
    \caption{The effect of aperture on the \lya\ profile for stacked subsets of the CLASSY sample. The median profile shapes are influenced by the effect of the optical to COS aperture size ($R_{50}/R_{COS}$) ratio, with larger values of $R_{50}/R_{COS}$ being associated with more substantial \lya\ absorption.}
    \label{fig:lymanalphaaperture}
\end{figure}

Finally, one more effect to consider is the influence of the spectroscopic aperture on the shape of the \lya\ profile. In Figure~\ref{fig:lymanalphaaperture}, we plot the median spectra at the intrinsic resolution of CLASSY galaxies binned according to the $R_{50}/R_{COS}$ ratio. $R_{50}$ is the optical half-light radius \citep[see][]{berg2022} and $R_{COS}$ the radius of the spectroscopic aperture. 
The Figure shows a clear trend of the \lya\ going from a pure emission profile for the smallest ratio, to a pure absorption profile for the largest ratio. Therefore, sources that are more extended than the spectroscopic aperture are seen as having greater \lya\ absorption. Typical spectroscopic surveys of EoR galaxies use NIRSpec in combination with the Micro Shutter Assembly (MSA), which have a typical size of 0\farcs1 $\times$ 0\farcs2, smaller than many of the EoR galaxies \citep{jakobsen2022}.

One more additional consideration is the fact that at high-z, it might be assumed that the metallicity of the universe should naturally decrease, 
and therefore the effect of the \nv\ line blending with \hi\ absorption may not be of substantial concern. However, given that the metallicities in the CLASSY mock spectra are similar to those found in the high-z universe \citep[e.g.,][]{williams2023,langeroodi2023}, and that unexpectedly large nitrogen enrichment is being found in high-z galaxies \citep[e.g.,][]{cameron2023,bunker2023,marqueschaves2024,topping2024}, the \nv\ line blending issue outlined in this work may actually be of quite significant concern in the high-z universe.

To conclude, although the use of the \hi\ damping wing imprinted on galaxy spectra is a promising avenue to constrain the history of the reionization of the universe, the uncertainty associated with this measurement still need to be carefully quantified.

\section{Conclusions}\label{sec:conclusion}
In this study, we use high spectral resolution data from the CLASSY galaxy survey to mock JWST NIRSpec Prism and grism  observations of galaxies during the EoR. 
The simulated spectra have known column densities of \hi\ reservoirs associated to each galaxy ($N_{\rm DLA}$) and known neutral hydrogen fraction ($x_{\rm HI}$) of the Universe. 
We use these spectra to assess the accuracy to which $x_{\rm HI}$ and $N_{\rm DLA}$ can be recovered in the JWST data. 
We find that there is a large scatter between the recovered and intrinsic values of the parameters. 
For most of  galaxies, the recovered column densities is within an order of magnitude from the intrinsic value, while the neutral hydrogen fraction is much less well constrained. 
We find that the recovered values of $N_{\rm DLA}$ and $x_{\rm HI}$ are strongly degenerate with each other and that the exact shape of the stellar continuum also introduces a potential source of degeneracy, particularly for the youngest galaxies. We find that increasing the spectral resolution results in some improvement of the recovered column densities but not for the $x_{\rm HI}$ fraction.

In Sections~\ref{sec:bubble} and~\ref{sec:bubbleshighres}, we expanded our simulations to include a bubble of ionized gas around the source galaxy, using both the low and high resolution simulated data. We find that the success in recovering  $N_{\rm DLA}$, and  $x_{\rm HI}$ improves marginally, compared to the case without the ionized bubble, more in the high resolution simulations. We note, however, that in real observations we will not know whether or not a bubble is present. Although detection of \lya\ in emission could potentially be used to infer the presence of a ionized bubble \citep{hayesscarlata2023}, we demonstrated in Section~\ref{sec:othereffects} that aperture effects make the use of \lya\ difficult to interpret. 

Finally we discussed in Section~\ref{sec:othereffects} physical effects that will introduce additional uncertainties in the recovered parameters, but that we did not explore in this work. Therefore, we recommend significant consideration of these sources of uncertainty and biases when using \lya\ damping wings in galaxies to measure the intergalactic neutral hydrogen fraction.

Our study reinforces the need for high spectral resolution deep infrared spectra, that will become available as the Extremely Large Telescope (ELT) commences operations by the end of this decade. The ELT Multi-Object Spectrograph (MOSAIC) \citep{kelz2015} and ELT ArmazoNes high Dispersion Echelle Spectrograph (ANDES) \citep{dimarcantonio2022} will have resolutions of R$>5000$ and $R\sim100000$ which will have the potential to drastically improve spectroscopy of damping wings in the distant universe. 

\section{Acknowledgments}
We thank Kasper Heintz for insightful feedback on this paper. We thank Danielle Berg and Peter Senchyna for providing Starburst99 fits for the original CLASSY spectra. We thank Weida Hu for providing CLASSY spectra with the Milky Way DLAs removed. M. Huberty thanks Alberto Saldana-Lopez for helpful discussion regarding the FICUS code. We acknowledge support by NASA through grant JWST-GO-1571. M.J.H. is supported by the Swedish Research Council (Vetenskapsr\aa{}det) and is Fellow of the Knut \& Alice Wallenberg Foundation.

\bibliography{refr}{}
\bibliographystyle{aasjournal}

\end{document}